\documentclass[aps,twocolumn]{revtex4}
\usepackage[dvips]{graphics,graphicx}
\usepackage{amsmath}
\begin{document}

\title{Bistability in the dissipative quantum systems I: \\ Damped and driven nonlinear oscillator }
\author{Andrey~R.~Kolovsky}
\affiliation{$^1$Kirensky Institute of Physics, 660036 Krasnoyarsk, Russia}
\affiliation{Siberian Federal University, 660041 Krasnoyarsk, Russia}
\begin{abstract}
We revisit  quantum dynamics of the damped and driven nonlinear oscillator. In the classical case this system  has two stationary solutions (the limit cycles) in the certain parameter region, which is the origin of the celebrated bistability phenomenon. The quantum-classical correspondence for the oscillator dynamics is discussed in details.  
\end{abstract}
\maketitle

\section{Introduction}

The damped and driven nonlinear oscillator is one of the paradigm model of classical physics which introduces and explains the phenomena of bistability and hysteresis \cite{Land76}. Thus, it not surprising that its quantum counterpart began to attract attention as early as in 1980 \cite{Drum80}. It was concluded in the cited paper that hysteresis and bistability are absent in the quantum approach. This challenging contradiction between the quantum and classical results was resolved by other researches \cite{Risk87,Voge88,Voge89,Voge90,Bort95,Rigo97} who identified the metastable character of the quantum limit cycles.  In the present work, which mainly follow the educational aims, we review the known results on the damped and driven nonlinear oscillator and complement them by analysis of the attractor basins and some results of the pseudoclassical approach. Being more involved than the classical analysis this approach is capable to reproduce the metastable character of the quantum limit cycles.  The presented studies create a platform for addressing more complex dissipative quantum systems, including those with chaotic dynamics \cite{preprint}. 

\section{Classical analysis}

\begin{figure}[h]
\includegraphics[width=8.5cm,clip]{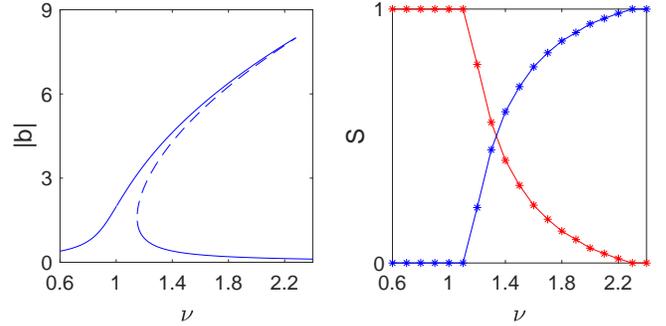}
\caption{Left panel: The stationary values of the oscillator amplitude as the function of the driving frequency. Right panel:  relative size of the basins of the outer (red line) and inner (blue line) limit cycles. The system parameters are are $\omega=1$, $g=0.02$,  $\epsilon=0.16$, and $\gamma=0.04$.}
\label{fig0}
\end{figure}

We are interested in the dynamics of the damped and driven nonlinear oscillator,
\begin{equation}
i\dot{a}=\frac{\partial H}{\partial a^*} -  i\frac{\gamma}{2} a \;,\quad  
\label{a0}
\end{equation}
\begin{equation} 
H= \omega a^*a + \frac{g}{2}(a^*a)^2 + \epsilon \left(e^{i\nu t}a +  e^{-i\nu t}a^* \right)  \;, 
\label{a1}
\end{equation}
where $\nu$ is the driving frequency, $g$ nonlinearity, and $\gamma$ the friction coefficient. For moderate negative detuning $\Delta \omega=\omega-\nu$  the system has two limit cycles,
\begin{equation}
a(t\rightarrow\infty)=b_{1,2} e^{-i\nu t} \;,
\label{a2}
\end{equation}
where $b_{1,2}$ are given by the stable solution of the following algebraic equation,
\begin{equation}
\left(\Delta\omega-i\frac{\gamma}{2}\right)b +g |b|^2b+\epsilon =0 \;.
\label{a3}
\end{equation}
Notice that Eq.~(\ref{a3}) can be recast into equations for the squared amplitude,
\begin{equation}
|b|^2=\frac{\epsilon^2}{(\Delta\omega+g|b|^2)^2+(\gamma/2)^2} \;,
\label{a4}
\end{equation}
and the phase,
\begin{equation}
\frac{{\rm Re}(b)}{|b|^2}=\frac{\Delta\omega+g|b|^2}{\epsilon} \;, \quad
\frac{{\rm Im}(b)}{|b|^2}=-\frac{\gamma}{2\epsilon}  \;.
\label{a5}
\end{equation}
Through the paper we use  $\omega=1$, $g=0.02$, $\gamma=0.04$, $\epsilon=0.16$, and the driving frequency  $\nu$, which is our control parameter, in the interval $0.6<\nu<2.4$.  For these parameters  solution of Eq.~(\ref{a4}) is depicted in the left panel in Fig.~\ref{fig0}. 

\begin{figure}[t]
\includegraphics[width=8.5cm,clip]{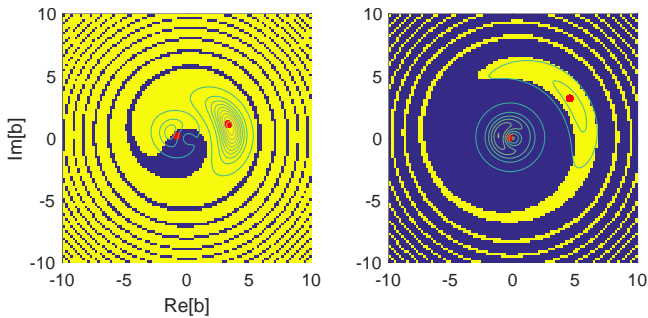}
\caption{Basins of the outer (yellow) and inner (blue) limit cycles for $\nu=1.2$ (left) and $\nu=1.6$ (right). The red dots depict solutions of Eq.~(\ref{a3}). Superimposed are contour lines of quantum attractors, see text.}
\label{fig00}
\end{figure}

Next we fix $\nu$ in the bistability region and find the basins of the attractors. Figure \ref{fig00} shows the basins of the limit cycles (red dots) for $\nu=1.2$ and $\nu=1.6$. It is seen that with  increase of the driving frequency the basin of the outer limit cycle gradually vanishes  and becomes zero for $\nu>\nu_{2}$ where Eq.~(\ref{a3}) has only one real solution. Analogously, for $\nu<\nu_1$ the inner cycle disappear.  For intermediate values of $\nu$ the relative size of the attractor basins (which we calculate by counting the blue and yellow pixels in the rectangular $-10\le {\rm Re}(a),{\rm Im}(a) \le 10$) is depicted in the right panel in Fig.~\ref{fig0}. We mention that these results allow us to introduce the third critical frequency $\nu_3$ where the basin sizes coincide.  

\begin{figure}
\includegraphics[width=8.5cm,clip]{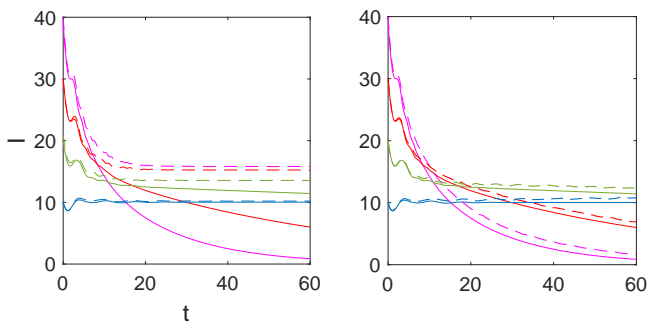}
\caption{The mean occupation number Eq.~(\ref{b4}), solid lines, as compared to the mean action Eq.~(\ref{a6}), dashed lines in the left panel, and the mean action Eq.~(\ref{d4}),  dashed lines in the right panel, for $\nu=1.2$, 1.4, 1.6, 1.8 (from bottom to top at $t=0$). The time is measured in units of $T=2\pi/\omega$.}
\label{fig1}
\end{figure}

Let us now address the system dynamics. As the initial condition we consider an ensemble of classical particles which are uniformly distributed  over a circle of the radius $a_0$, i.e., $a(t=0)=a_0e^{i\theta}$, $0\le \theta <2\pi$.  Then  the relative number of particles attracted to the outer and inner cycles is determined by the relative size of  the `yellow'  and `blue' segments of the circle.  For the sake of future comparison  the dashed lines in the left panel in Fig.~\ref{fig1} show dynamics of the mean action $I(t)$,
\begin{equation}
I(t)=\overline{|a(t)|^2} \;,
\label{a6}
\end{equation}
as the function of time for $\nu=1.2,1.4,1.6$ and $a_0=|b_2(\nu)|$. Relaxation to the stationary regime within the characteristic time $T_\gamma=2\pi/\gamma$ is clearly seen.
%

\section{Quantum analysis}

Setting the Planck constant to unity, the master equation for the damped and driven nonlinear oscillator reads
\begin{equation}
\label{b0}
\frac{d \hat{\rho}}{dt}=-i[\widehat{H},\hat{\rho}]  + \widehat{{\cal G}}(\hat{\rho}) \;,
\end{equation}
where
\begin{equation}
\label{b1}
\widehat{H}=\omega \hat{a}^\dagger \hat{a} + \frac{g}{2} (\hat{a}^\dagger \hat{a})^2 +\epsilon\left(e^{-i\nu t} \hat{a}^\dagger + h.c.\right)   
\end{equation}
is the system Hamiltonian,  and $\widehat{{\cal G}}(\hat{\rho})$,
\begin{equation}
\label{b2}
\widehat{{\cal G}}(\hat{\rho})=-\frac{\gamma}{2}
 (\hat{a}\hat{a}^\dagger\hat{\rho}-2\hat{a}^\dagger\hat{\rho}\hat{a} + \hat{\rho}\hat{a}\hat{a}^\dagger) \;,
\end{equation}
is the Lindblad relaxation operator.  The structure of this operator is fixed by the condition that in the absence of driving the oscillator relaxes into the ground state.

\begin{figure}
\includegraphics[width=8.5cm,clip]{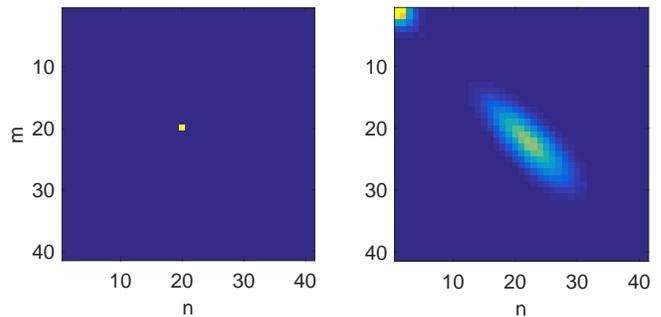}
\caption{The density matrix in the energy basis at $t=0$, left, $t=20T$, right. The system parameters are $\omega=1$, $g=0.02$, $\epsilon=0.16$, $\gamma=0.04$, and $\nu=1.4$.}
\label{fig3}
\end{figure}

We solve the master equation (\ref{b0}) in the energy basis for the initial condition that corresponds to population of the single energy level with the index $n_0$. As an example the left panel in Fig.~\ref{fig3} shows  the matrix elements $\rho_{n,m}(t)$ at $t=20T$ for $\nu=1.4$.  Signatures of the classical limit cycles are clearly seen. 
\begin{figure}
\includegraphics[width=8.8cm,clip]{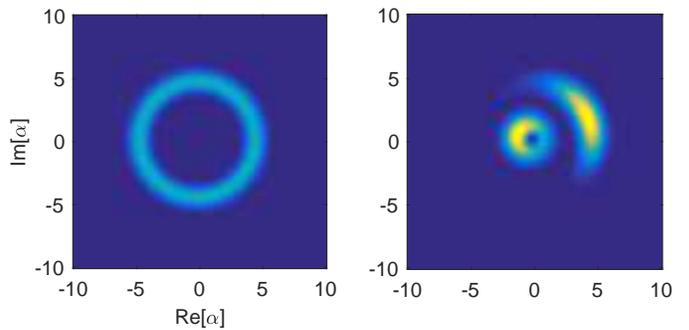}
\caption{The phase-space representation of the density matrix shown in the previous figure.}
\label{fig4}
\end{figure}

The solid lines in Fig.~\ref{fig0} show the mean occupation number $n(t)$,
\begin{equation}
n(t)={\rm Tr}[\hat{a}^\dagger\hat{a} \hat{\rho}(t)] \;,
\label{b4}
\end{equation}
as the function of time for the same values of the parameters which were used to calculate the classical dynamics. For $\nu=1.2$ we observe a good agreement with the classical result while for a larger $\nu$ there are considerable deviations.   To understand  the origin of these deviations on the qualitative level we consider the density matrix in the phase-space representation, 
\begin{equation}
\rho(\alpha,t)=\langle \alpha |\hat{\rho}(t) |\alpha \rangle \;,
\label{b5}
\end{equation}
where $|\alpha\rangle$,
\begin{equation}
|\alpha\rangle = \exp\left(\frac{|\alpha|^2}{2}\right) \sum_{n=0}^\infty \frac{\alpha^n}{\sqrt{n!}} |n\rangle   \;, 
\label{b6}
\end{equation}
is the coherent Glauber state.  The initial and final density matrices in the phase-space representation are shown in  Fig.~\ref{fig4}.  As expected, the quantum attractors are not the $\delta$-functions but distributions with a finite width.  In Fig.~\ref{fig00} we projected the contour lines of these distributions on the basins of the classical attractors. It is seen that for $\nu=1.2$ the region of support of the outer quantum attractor is well inside the basin of the classical attractor. As it will be explained in Sec.V, this results in an exponentially long lifetime of the outer (quantum) limit cycle and, as  the consequence, in a good agreement between the classical and quantum dynamics. Unlike the case $\nu=1.2$, for $\nu=1.4$ the tails of the outer quantum attractor go outside the basin of the classical attractor that results in considerably shorter lifetime of the outer limit cycle.

\section{Lifetime of the limit cycle}

To quantify lifetimes of the limit cycles we rewrite the master equation (\ref{b0}) in the form
\begin{equation}
\frac{{\rm d}\hat{\rho}}{{\rm d} t}=\widehat{L}\hat{\rho} \;,
\label{c1}
\end{equation}
where the linear operator $\widehat{L}$ is often referred to as the super-operator. Due to linearity  of Eq.~(\ref{c1}) its solution has the form
\begin{equation}
\hat{\rho}(t)=\sum_{j=0}^\infty \hat{\rho}^{(j)} e^{\lambda_j t} \;, \quad |\lambda_j|\le |\lambda_{j+1}| \;,
\label{c2}
\end{equation}
where $\lambda_j$ and $\hat{\rho}^{(j)}$ are eigenvalues and `eigen-matrices' of the super-operator $\widehat{L}$. Notice that the real parts of $\lambda_j$ are negative except for $\lambda_0$ which is strictly zero. This ensures relaxation of the density matrix $\hat{\rho}(t)$ into the steady state $\hat{\rho}^{(0)}$ within the characteristic relaxation time
\begin{equation}
\tau=\frac{2\pi}{|{\rm Re}(\lambda_1)| } \;.
\label{c3}
\end{equation}
Numerically one finds $\lambda_j$  by truncating the density matrix  to a finite size $N\times N$ and constructing the column vector ${\cal R}$ of the length $N^2$ by re-ordering the matrix elements $\rho_{n,m}$ in the column-wise manner. Then the super-operator  $\widehat{L}$ is given by a matrix $L$ of the size $N^2 \times N^2$ and the matrices   $\hat{\rho}^{(j)}$ are obtained  by re-ordering the eigenvectors  ${\cal R}^{(j)}$ of this matrix  back to the  $N\times N$ square matrices. 
\begin{figure}
\includegraphics[width=8.5cm,clip]{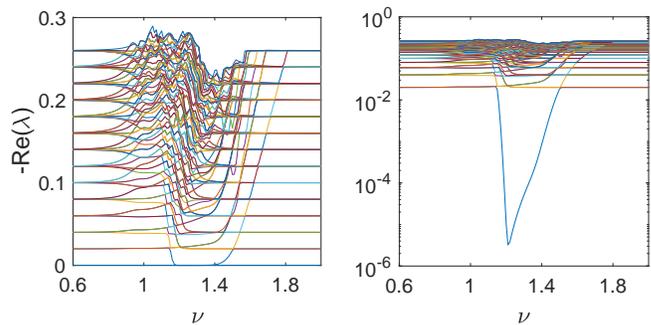}
\caption{The first 100 smallest eigenvalues of the super-operator $\widehat{L}$ as the function of $\nu$ in the linear and logarithmic scales.}
\label{fig7}
\end{figure} 
\begin{figure}
\includegraphics[width=8.5cm,clip]{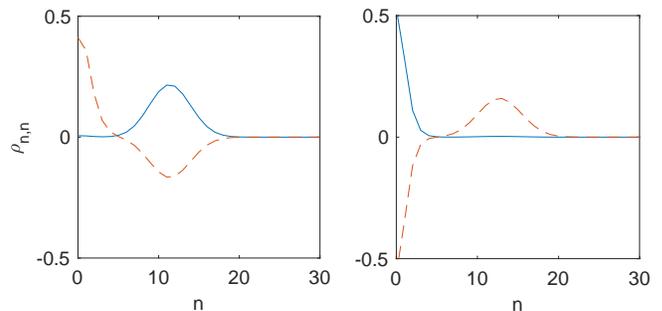}
\caption{Diagonal elements of the matrices $\hat{\rho}^{(0)}$ (solid line) and  $\hat{\rho}^{(1)}$ (dashed line) for $\nu=1.19$, left panel, and $\nu=1.22$, right panel.}
\label{fig8}
\end{figure}

Figure \ref{fig7} shows the real parts of the first 100 eigenvalues $\lambda_j$  as the function of the detuning $\nu$. One easily identifies in the depicted level pattern the decay spectrum of detuned harmonic oscillator,
\begin{equation}
{\rm Re}(\lambda_j) =-\frac{\gamma}{2} j  \;,
\label{c4}
\end{equation}
and the eigenvalue associated with the metastable limit cycle.  The frequency interval where ${\rm Re}(\lambda_1)<\gamma/2$ is the quantum bistability region where the system has one stable and one metastable attractors. Which of two attractors is metastable is determined by the inequality relation  between $\nu$ and $\nu_3\approx 1.2$ where $|{\rm Re}(\lambda_1)|$ is minimal. We also mention that in the quantum bistability region  the long-time dynamics of the system  is determined by the equation
\begin{equation}
\hat{\rho}(t)=\hat{\rho}^{(0)} + \exp(-t/\tau) \hat{\rho}^{(1)}  \;.
\label{c5}
\end{equation}
Figure \ref{fig8} shows the diagonal elements of the matrices $\hat{\rho}^{(0)}$ and  $\hat{\rho}^{(1)}$ for $\nu=1.19<\nu_3$, left panel, and $\nu=1.22>\nu_3$, right panel.  (Notice that ${\rm Tr}[\hat{\rho}^{(0)}]=1$ whereas ${\rm Tr}[\hat{\rho}^{(1)}]=0$.) Thus, in the former case Eq.~(\ref{c5}) describes probability leakage from the inner into the outer limit cycle, while in the latter case the situation is inverted. In both cases, however,  lifetime of the metastable cycle exceeds the classical relaxation time $T_\gamma$ by orders of magnitude.

\section{Pseudo-classical approach}

In this section we show that the metastable character of quantum limit cycles can be well reproduced by using the pseudo-classical approach. The starting point of this approach is the equation on the Wigner function  $w=w(a,a^*,t)$ of the quantum oscillator which is uniquely determined by the system density matrix $\hat{\rho}(t)$ and is a real function of two complex variables $a$ and $a^*$.  Applying the Weyl transform to the master equation (\ref{b0}) the equation on the Wigner function reads
\begin{eqnarray}
\label{d1}
\frac{\partial w}{\partial t}
=\{ H,w\}  
-i\frac{g}{4}\left( a\frac{\partial^3 w}{\partial^2 a \partial^2 a^*} - a^*\frac{\partial^3 w}{\partial^2 a^* \partial a} \right) \\
\nonumber
+\frac{\gamma}{2}\left( a\frac{\partial w}{\partial a} +2w+ a^*\frac{\partial w}{\partial a^*} \right) 
+\frac{\gamma}{2} \frac{\partial^2 w}{\partial a \partial a^*} \;,
\end{eqnarray}
where $H$ is the classical Hamiltonian (\ref{a1}) and $\{\ldots,\ldots \}$ denotes the Poisson brackets. In this equation the first line corresponds to the unitary evolution of the system and the last line is the Weyl image of the Lindblad relaxation operator.  The pseudo-classical approximation (which is also known as the truncated Wigner function approximation) amounts to neglecting all terms  which involve higher than second derivative. Then Eq.~(\ref{d1}) takes the form of the Fokker-Planck equation on the classical distribution function $f=f(a,a^*,t)$,
\begin{eqnarray}
\label{d2}
\frac{\partial f}{\partial t}=\{ H,f\}  \\
\nonumber
+\frac{\gamma}{2}\left( a\frac{\partial f}{\partial a} +2f+ a^*\frac{\partial f}{\partial a^*} \right) 
+\frac{\gamma}{2} \frac{\partial^2 f}{\partial a \partial a^*} \;,
\end{eqnarray}
By physical meaning the first term in the r.h.s. of Eq.(\ref{d2}) is the Hamiltonian evolution of the system, the second term is the friction term responsible for the phase space contraction, and the last term is the diffusion term due to irreducible quantum noise on which we briefly comment in the next paragraph. 

The origin of the term `irreducible quantum noise' becomes clear if one consider a slightly more complex problem where the relaxation term in the master equation has the form 
\begin{eqnarray}
\label{d3}
\widehat{{\cal G}}(\hat{\rho})=
-\frac{\gamma(\bar{n}+1)}{2} (\hat{a}\hat{a}^\dagger\hat{\rho}-2\hat{a}^\dagger\hat{\rho}\hat{a} + \hat{\rho}\hat{a}\hat{a}^\dagger) \\
\nonumber
-\frac{\gamma\bar{n}}{2} (\hat{a}^\dagger\hat{a}\hat{\rho}-2\hat{a}\hat{\rho}\hat{a}^\dagger + \hat{\rho}\hat{a}^\dagger\hat{a}) \;,
\end{eqnarray}
In the absence  of driving this operator causes relaxation of the quantum oscillator not in the ground state but into the Boltzman state where $\rho_{n,n} \sim \exp(-n/\bar{n})$. For this problem the pseudoclassical approach gives the same Fokker-Planck equation where, however, the diffusion term is preceded by the  coefficient $\gamma(2\bar{n}+1)/2$. Thus, even if we have relaxation to the ground state (i.e., if we set $\bar{n}=0$), we still have the diffusion term. 

To solve the Fokker-Planck equation (\ref{d2})  we employ the Monte-Carlo approach. In fact, it is easy to show that Eq.~(\ref{d2}) is equivalent to the following Langevin equation on the canonical variable $a(t)$,
\begin{equation}
\label{d4}
i\dot{a}=\frac{\partial H}{\partial a^*} -  i\frac{\gamma}{2} a + \sqrt{\frac{\gamma}{4}} \xi(t) \;.
\end{equation}
Here $\xi(t)$ is the complex white noise with $\xi^*(t)\xi(t')=2\delta(t-t')$.  Using Eq.~(\ref{d3}) we calculate dynamics of the mean action, 
$I(t)=\langle \overline{|a(t)|^2} \rangle$,
%
where, as before, the bar denotes the average over ensemble of initial conditions  and the angular brackets are additional average over different  realizations of the random process $\xi(t)$. It is seen in Fig.~\ref{fig1}(b) that pseudo-classical approach well reproduces the quantum dynamics of the system.

%


\section{Conclusion}

We analyzed the quantum damped and driven nonlinear oscillator in the bistable region where its classical counterpart has two limit cycles.  
It was shown that these limit cycles are seen in the quantum dynamics of  the oscillator as well.  However,  unlike the classical case, one of these cycles has finite lifetime which may vary from a fraction of the relaxation period  $T_\gamma$ to thousands relaxation periods. Remarkably, this lifetime can be well estimated by using the pseudoclassical approach which takes into account the irreducible quantum noise with the intensity $\gamma\hbar/2$. (Here we use the unscaled variables, i.e., the fundamental Planck constant is not set to unity). This random force can kick the system out the basin of one attractor into the basin of other attractor that results in a gradual depopulation of the attractor with smaller basin size. We mention with this respect that it is very important to analyze the basins of attractors. Unfortunately, this analysis was missing in Ref.~\cite{Risk87,Voge88,Voge89,Voge90,Bort95,Rigo97}  that prevented  the authors of the cited works from establishing the full quantum-classical  correspondence in the considered paradigm model of quantum bistability.



\begin{thebibliography}{99}


\bibitem{Land76}
L. D. Landau and E. M. Lifshitz, 
{\em Mechanics} (Pergamon, New York, 1976).

\bibitem{Drum80}
P.D. Drummond and D.F. Walls, 
{\em Quantum theory of optical bistability. I. Nonlinear polarisability model.}
J. Phys. A {\bf 13}, 725 (1980).

\bibitem{Risk87}
H. Risken, C. Savage, F. Haake, and D. F. Walls,
{\em Quantum tunneling in dispersive optical bistability},
Phys. Rev. A {\bf 35}, 1729 (1987).

\bibitem{Voge88}
K. Vogel and H. Risken, 
{\em Quantum-tunneling rates and stationary solutions in dispersive optical bistability},
Phys. Rev. A {\bf 38}, 2409 (1988).

\bibitem{Voge89}
K. Vogel and H. Risken, 
{\em Quasiprobability distributions in dispersive optical bistability},
Phys. Rev. A {\bf 39}, 4675 (1989).

\bibitem{Voge90}
K. Vogel and H. Risken,
{\em Dispersive optical bistability for large photon numbers and low cavity damping}
Phys. Rev. A {\bf 42}, 627 (1990).

\bibitem{Bort95}
D. Bortman and A. Ron, 
{\em Bistability in a quantum nonlinear oscillator},
Phys. Rev. A {\bf 52}, 3316 (1995).

\bibitem{Rigo97}
M. Rigo, G. Alber, F. Mota-Furtado, and P. F. O'Mahony,
{\em Quantum-state diffusion model and the driven damped nonlinear oscillator},
Phys. Rev. A {\bf 55}, 1665 (1997).

\bibitem{preprint}
A. R. Kolovsky,
{\em Bistability and chaos assisted tunnelling in the dissipative quantum systems},
in progress.

\end{thebibliography}
\end{document}